# Development, Calibration and Characterization of Silicon Photonics based Optical Phased Arrays

Sylvain Guerber[a], Daivid Fowler[a], Ismael Charlet[a,b], Philippe Grosse[a], Kim Abdoul-Carime[a], Jonathan Faugier-Tovar[a], and Bertrand Szelag[a]
[a] CEA, LETI, MINATEC Campus, CEA-Grenoble, F-38054 Grenoble, France
[b] STMicroelectronics, 850 rue Jean Monnet, 38920 Crolles, France

## ABSTRACT

Over the last decade, Optical Phased Arrays (OPA) have been extensively studied, targeting applications such as Light Detection And Ranging (LiDAR) systems, holographic displays, atmospheric monitoring and free space communications. Leveraging the maturity of the silicon photonics platform, the usual mechanical based beam steering system could be replaced by an integrated OPA; significantly reducing the cost and size of the LiDAR while improving its performance (scanning speed, power efficiency, resolution…) thanks to solid state beam steering. However, the realization of an OPA that meets the specifications of a LiDAR system (low divergence and single output beam) is not trivial. Targeting the realization of a complete LiDAR system, the technical challenges inherent to the development of high performance OPAs have been studied at CEA LETI. In particular, efficient genetic algorithms have been developed for the calibration of high channel count OPAs as well as an advanced measurement setup compatible with wafer-scale OPA characterization.

## 1. INTRODUCTION

Driven by the massive and imminent deployment of autonomous cars, LiDAR (Light Detection And Ranging) systems have been the subject of growing interest over the last few years. Indeed, compared to other 3D imaging technologies such as RADAR or cameras, LiDAR systems can provide a high detection range, finer angular resolution and much-reduced sensitivity to ambient conditions. However, current LiDAR sensors generally rely on heavy, slow, power hungry and expensive mechanical beam steering mechanisms, preventing the large-scale deployment of this technology. Recently, another approach based on optical phased arrays (OPA) has been proposed [1-4]. Taking benefits from the maturity of the silicon photonics platform, the usual mechanical-based beam steering system could be replaced by an integrated chip-scale OPA. This would significantly reduce the physical size and cost of the LiDAR system while improving its performances (scanning speed, power efficiency, resolution…) thanks to solid state beam steering. Thus, the development of a high performance OPA would pave the way to cheap LiDAR systems for automotive applications [5] as well as free space communications [6], holographic displays [7] and biomedical imaging [8].

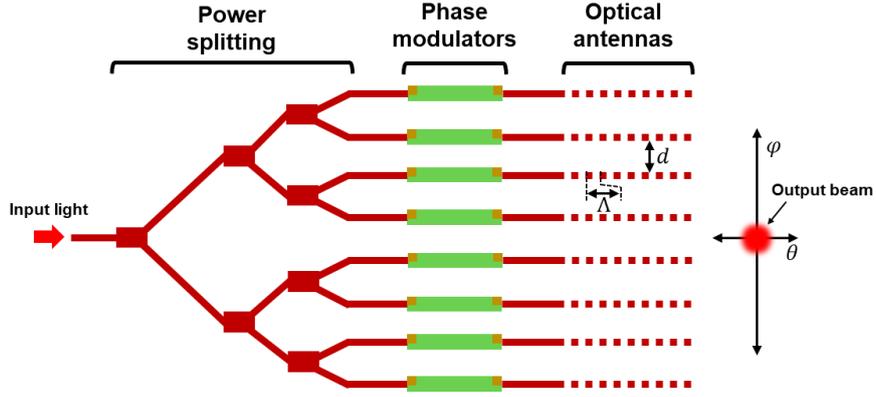

Figure 1, Schematic view of an OPA.

A schematic view of an OPA is presented Figure 1. Firstly, laser light is coupled in a single waveguide of the photonic chip. Then, using a series of multimode interferometers (MMI) arranged as a splitting tree, the available optical power is (ideally) equally distributed between many waveguides. Each waveguide is then connected to an optical antenna which consists of a diffraction grating. The output interference pattern will depend on the phase relation between the antennas which can be tuned thanks to phase modulators.

One of the important parameter of a high resolution LiDAR system is the divergence $\varphi_{3dB}$ (°) of the produced output beam which should be small enough to satisfy the resolution requirements of the system. In the case of an OPA based LiDAR, this is defined by the operating wavelength λ (nm), the number N and pitch d (nm) of optical antennas as stated in equation (1).

$$\varphi_{3dB} = \frac{70\lambda}{Nd} \qquad (1)$$

Since a large optical aperture is required to produce a low divergence beam, a high channel count and large antenna spacing would be desirable. However, assuming an OPA with uniform channel spacing, the interference pattern of the emitted light will present several diffraction orders whose angular distance will depend on the antenna pitch d as stated in equation (2).

$$\sin(\Delta\varphi) = \frac{\lambda}{d} \qquad (2)$$

Therefore, in order to remove those unwanted high order lobes which limit the OPA unambiguous field of view (FOV), the antennas pitch should be kept small. OPAs with aperiodic antennas pitch have been proposed to suppress the high order lobes while maintaining large antenna spacing [9]. While this indeed allows a greater FOV, for the same average antenna spacing, the main lobe power fraction does not improve, as the power fraction normally contained in the higher order lobes is distributed evenly over the noise floor. Thus, the inherently large optical losses makes this approach not compatible with long range LiDAR which requires a large amount of optical power in the main beam. As summarized in Figure 2, to have a low divergence beam and wide field of view, an OPA must have a large amount of closely spaced antennas. As a consequence, the number of optical antennas in a single OPA has been significantly increased over the years [4].

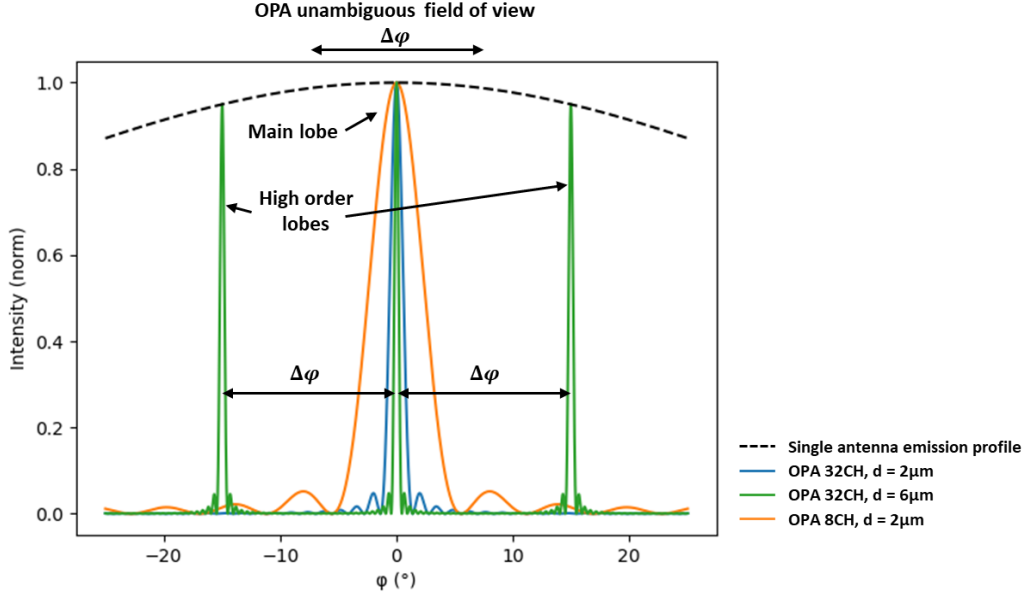

Figure 2, Analytical simulation of OPA with different channel and antenna pitch configurations (λ=1550nm).

As mentioned above, the emission profile of an OPA depend on the phase relation between the optical antennas. In order to produce an optimally shaped beam at a specific angle in the φ direction, a linear phase gradient ψ must be applied between the antennas as stated in equation (3).

$$\sin(\varphi) = \frac{\psi\lambda}{2\pi d} \qquad (3)$$

Therefore, considering an OPA designed with equal waveguide lengths, a centered (φ = 0°) and nicely shaped output spot should be obtained, with no applied phase modulation. Yet, due to waveguide imperfections inherent in the fabrication process (surface roughness, dimension variation…), the designed and fabricated optical length of each OPA channel will differ, resulting in a low quality output beam. Fortunately, the phase modulators meant for beam steering can be used to compensate for these phase errors. However, due the random nature of the phase errors, all of the OPA channels must be calibrated. In an OPA with a large amount of optical channels, an efficient algorithm must be used to reduce the duration of the OPA calibration Furthermore, in order to facilitate OPA development and manufacture, this calibration phase must be done quickly and at the wafer level

The manuscript is organized as follow. In section 2 we present a comparison between two calibration algorithms. In section 3 we present our wafer level OPA characterization setup. Then, the proposed algorithm and characterization setup have been used to drive a 16-channel OPA, those results are presented in section 4. Finally, a brief discussion and conclusion are presented in section 5.

## 2. OPA CALIBRATION

In order to perform the calibration of an OPA, two principle elements are required: an optimization algorithm and a figure of merit (FOM) that should be maximized (or minimized). In this study, the FOM is defined as the overlap integral between the OPA far field and a Gaussian beam centered at the specified output angle as presented Figure 3(a). The most commonly used algorithm for OPA calibration is probably of the hill climbing type [10]. The principle of this deterministic algorithm is presented on Figure 3(b). Selecting one optical channel a time, the channel phase will be varied progressively between 0 and 2π. For each possible phase value, the FOM is computed. Once all values have been tested, the one that provides the highest FOM is selected and the algorithm switch to the next antenna until all antennas have been optimized. This sequence may be repeated a few times to allow the FOM to converge. The hill climbing algorithm is relatively easy to implement and to use since there are no tunable parameters. However, as detailed below in this section, the calibration duration does not scale well as the number of channels and number of phase discretization levels increase. Therefore, for high channel count OPAs, another option has been explored.

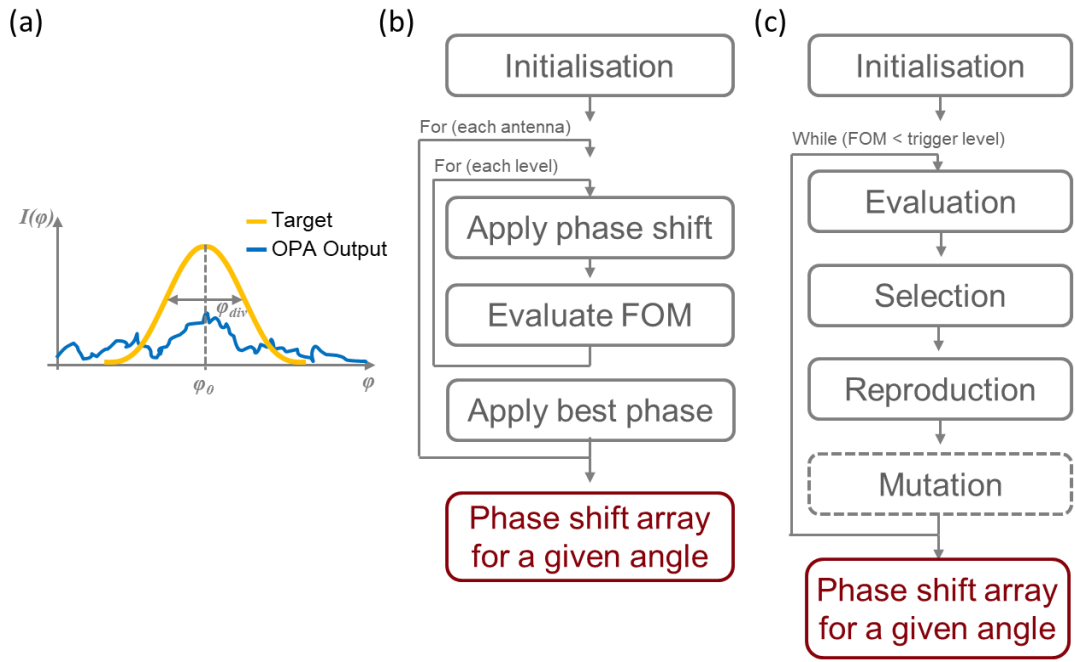

Figure 3, (a) Illustration of the OPA output target for FOM calculation. Flow of the (b) hill climbing and (c) genetic algorithm.

The working principle of genetic algorithms is directly inspired by Charles Darwin's law of evolution. A summary is presented Figure 3(c). First, a set of individuals which are each constituted of N genes (N the number of channels of the OPA) is generated during the initialization. Each gene represents the phase of an OPA channel. The FOM is then computed for each individual (evaluation). Then, a set of individuals is selected based on their FOM. Several selection mechanism can be used:

- Ranking: The individual with the best FOM is selected, then the second … until a certain amount (selection rate) of individual is selected.

- Tournament: Two individuals are selected randomly and the one with the best FOM is kept. This process is repeated until a certain amount (selection rate) of individual is selected.

- Fortune wheel: The individual are represented on a wheel with an angular part as large as their FOM. An angle is selected randomly and the corresponding individual is selected. The larger the FOM the higher the selection chance. This process is repeated until a certain amount (selection rate) of individual is selected.

Depending on the problem to solve, either one method or another may work better. In our case, only the ranking method has been tested. Once a set of individuals has been selected, the reproduction takes place. Two individuals (parents) are selected and a new individual is generated. Each gene of the new individual (child) is taken randomly from one of the parents. During this process mutation may occur. For a given mutation rate, some of the child's genes will be slightly modified. The novel gene value is randomly selected around its initial value within a certain range (mutation force). This process is repeated until the population size is back to its initial level (before selection). Then, the whole process (evaluation, selection, reproduction, mutation) is repeated on this novel generation. Compared to hill climbing, genetic algorithms are much more difficult to implement due to the large number of parameters (population size, selection rate, mutation rate, mutation force) and more complicated to use. However as detailed below, if well calibrated, they can be very powerful.

Among other indicators, the convergence speed of an algorithm is an important metric. It is measured as the number of evaluations of the FOM that is necessary to find an optimum value or reach a certain level of convergence. In order to evaluate the convergence speed of the proposed calibration algorithms, an analytical model based on the 1D grating model

[11] has been developed. Taking the OPA parameters as inputs (Δφ, N, d, λ, single antenna emission profile), this model can quickly compute the theoretical of the OPA-far field. Such a model has been used to test and develop our OPA calibration algorithms. An example is shown Figure 2. In an OPA with random initial phase values, both algorithms have been run with a target defined as a Gaussian centered at 0°. The far field of the OPA is computed using the aforementioned analytical model, these results are presented Figure 4. To begin, we considered an OPA with 256 antennas whose phase modulators are driven by 4 bit DACs which correspond to 16 phase discretization level between 0 and $2\pi$. As it can be seen on Figure 4(a), for this configuration both algorithms show quite similar convergence speeds. The discontinuity around the 4096th (256*16) evaluation for the hill climbing corresponds to the end of the first optimization cycle (each channel has been optimized once), two full optimization cycles have been run here. However, if the number of discretization levels is increased, to 256 (8 bit DAC) Figure 4(b), or to 4095 (12 bit DAC) Figure 4(c), a large discrepancy in convergence speed between the two algorithms appears. This is due to the deterministic nature of the hill climbing. Indeed, this algorithm is systematically exploring all possible phase values for each channel, which is extremely time-consuming. Thus, an increase in discretization level leads to a much greater number of evaluations of the FOM. Obviously, the same behavior is observed if the number of channels is increased. On the contrary, the genetic algorithm is much less sensitive to the increased size of the solution space because it spends less time evaluating sub-optimal solutions. Therefore, in the case of an OPA with a large number of channels, genetic algorithms can achieve far greater convergence speed. Moreover, the genetic algorithm is much more reliable in finding the optimal solution. Figure 4(d) presents the optimization of a 128 channel OPA with 128 levels of phase discretization. Each algorithm has been run five times with a different initial random phase distribution. As can be seen, the genetic algorithm finds a relatively similar FOM value. As a deterministic algorithm, the hill climbing is extremely dependent on the initial phase distribution. Therefore, the optimal value that is found varies significantly from one optimization to another.

Thus, the genetic algorithm seems to perform very well for OPA calibration, it has been used to calibrate the 16CH Silicon based OPA presented in section 5 of this manuscript.

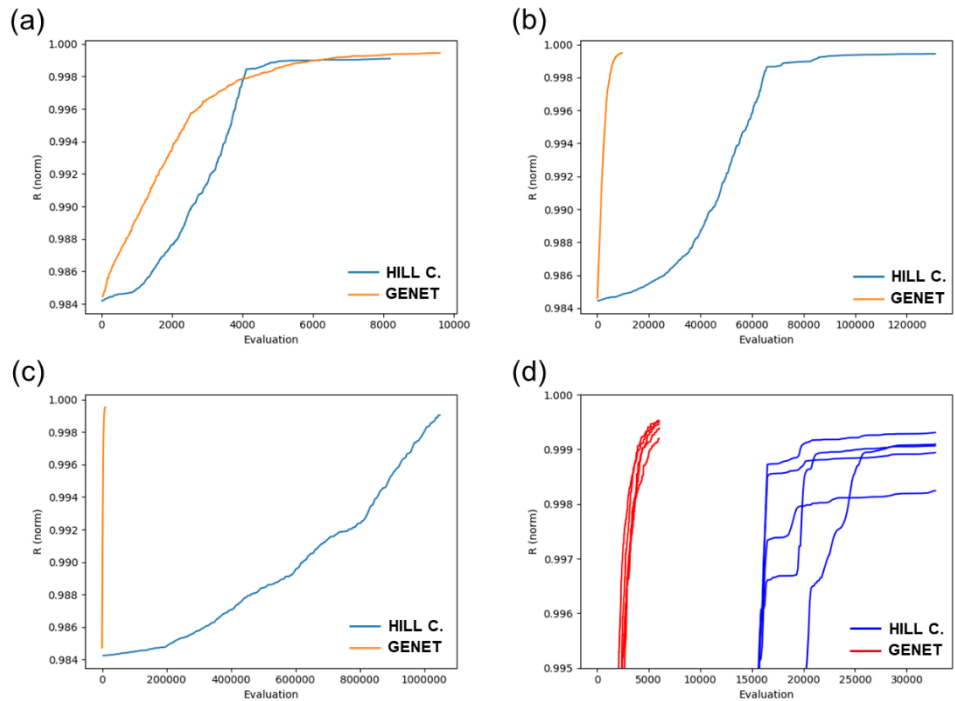

Figure 4, Evolution of the figure of merit R with respect to the number of evaluations for a 256CH OPA with (a) 16, (b) 256, (c) 4096 phase discretization levels for both hill climbing and genetic algorithms. Comparison of the convergence efficiency of hill climbing and genetic algorithm for a 128CH OPA with 128 phase discretization levels.

## 3. CHARACTERIZATION SETUP

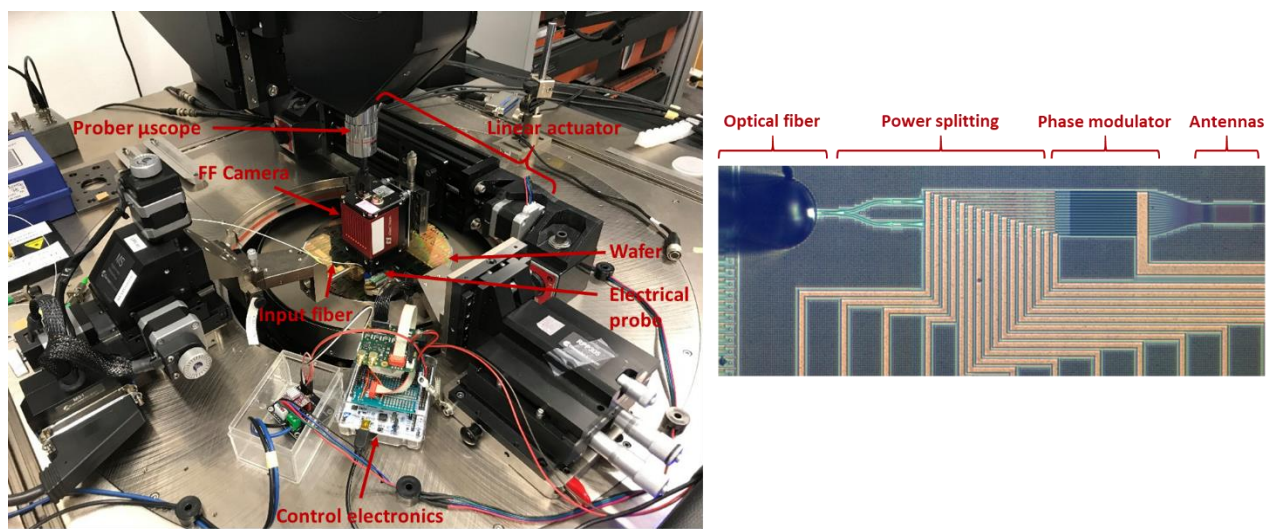

Figure 5, Overview of the wafer level OPA characterization bench. Microscope view of a 16CH Si based OPA.

As presented in the previous section, the calibration of an OPA is not trivial and may take a lot of time. In fact, multiple calibration cycles will be required on a single OPA, since each angular value will need dedicated optimization. Moreover, a calibration that is performed at ambient temperature may not be valid if the temperature changes since it will affect the phase of light propagating within the OPA waveguides. Therefore, OPAs targeting automotive applications may require multiple calibration tables depending on the LiDAR temperature operating range and OPA temperature sensitivity.

Thus, production of an OPA based LiDAR will require a fast and efficient characterization system. Traditionally, high speed testing is done on electronic and photonic devices at wafer level. Automatic probing stations (prober) are used to test multiple circuits on the different dies of a wafer. In this context, we developed a wafer level OPA characterization bench based on a probing station, a picture is presented Figure 5. The photonic wafer (up to 300mm diameter) is loaded in the probe station. An optical fiber is aligned with the input grating coupler of the OPA in order to inject laser light in the circuit. A multi-channel electric probe is used to contact simultaneously all the phase modulators of the OPA. In order to characterize, calibrate and drive the OPA, its far field must been imaged. Usually, this is done using a camera and a set of optical lenses [10] allowing for a large field of view. However, due to the lack of available space on the probing station, the use of a large optical system is not possible. Therefore, we choose to perform direct imaging of the OPA far field directly onto the focal plane of a camera [12, 13]. This works well, however the field of view of this setup is limited to a few degrees due to the small camera sensor size and minimum distance between the wafer and the camera (a few cm). To overcome this limitation, we introduce a linear actuator that can slide the camera along the φ direction as presented Figure 5. The displacement is accurately controlled by a worm drive and stepper motor. Using this system, the camera can be centered at any angle in the φ direction. Moreover, as can be seen in Figure 6(a), large field (panoramic) images can be obtained by moving the camera with steps equal to the sensor size and stitching the images.

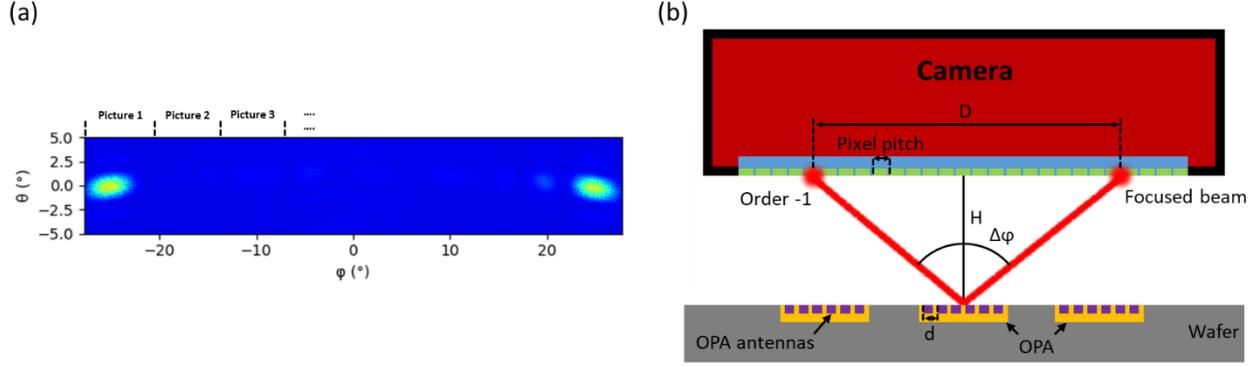

Figure 6, (a) Panoramic image of the far field of an OPA. (b) Principle of the angular calibration of the bench.

In order to have an angular scale on the camera (°/pixel), the setup must be calibrated. The procedure is described Figure 6(b). First, a rough estimation of the angular scale is performed using camera pixel pitch and by measuring manually the distance between the camera and the wafer surface, H. Then, the angular distance between two adjacent diffraction orders of the OPA is computed based on the OPA parameters and equation (2). For an OPA with an antenna pitch of 2µm and working at 1550nm we have $\Delta\varphi = 50.8°$. The OPA is then calibrated to produce a beam at $\Delta\varphi/2$. As shown on Figure 6(a), a second order lobe is produced at - $\Delta\varphi/2$. Then, the number of pixels in between the two lobes is measured by fitting the beams with Gaussian curves. Thus, a precise angular scale (°/pixel) is obtained. However, the absolute vertical angular origin must still be determined since this process provides only relative angles. Assuming that the beam with the highest power should be obtained for $\varphi = 0°$, the absolute origin car be found by doing multiple calibration at different angles while monitoring the beam power. Another possibility is to find the angle $\varphi$ for which the power of the main lobe and second order lobe is equalized. Thus, the absolute origin is in between the two beams.

## 4. 16CH OPA CARACTERISATION

As a demonstration, the calibration algorithm and characterization bench have been used to drive a silicon based 16CH OPA operating at $\lambda = 1550$nm and with an antenna pitch of 2µm. To begin, a test structure consisting of a waveguide connected to an antenna has been characterized in order to measure the emission profile of a single antenna (array factor). The result is presented on Figure 7(a). By plotting the profile in 1D as presented in Figure 7(b), two side lobes in the emission profile corresponding to a sinc function can be seen. This will allow for an accurate calibration of the OPA since the maximum optical power achievable in a beam for a specific angle is determined by this envelope function.

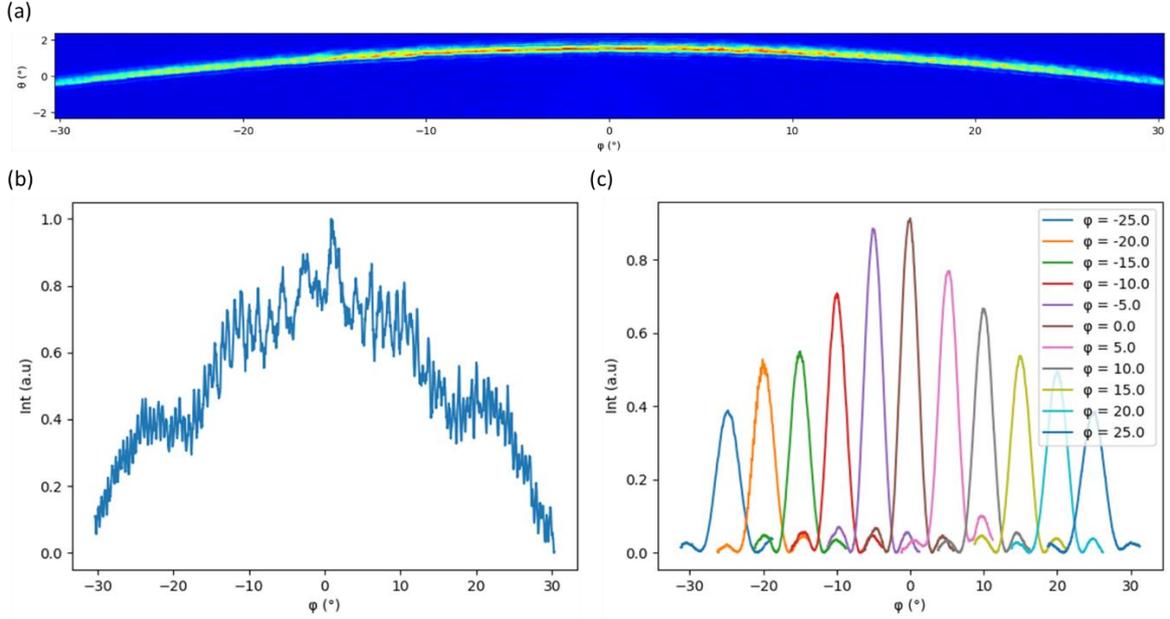

Figure 7, (a) Panoramic image of the emission profile of a single antenna in (a) 2D and (b) 1D. (c) Calibration of the 16 channel OPA between -25° and +25° with 5° steps.

The OPA can now be calibrated. Using the linear actuator, the camera is centered to the desired output beam angle $\varphi$. Then, using the developed genetic algorithm, the beam is optimized according to the given figure of merit defined as a gaussian centered at $\varphi$. This process is repeated for each desired output angle of the OPA. As shown Figure 7(c), the OPA has been calibrated between -25 and +25° with 5° steps. As expected, the beam envelope follows the single antenna emission profile.

## 5. CONCLUSION

Leveraging the maturity of the silicon photonics platform, the usual mechanical based beam steering system could be replaced by an integrated OPA; significantly reducing the cost and size of a LiDAR while improving performance (scanning speed, power efficiency, resolution…) thanks to solid state beam steering. However, the realization of an OPA that meet the specifications of a LiDAR system (low divergence and single output beam) is not trivial. Moreover, in order to achieve large scale production of OPA based LiDAR, an efficient characterization system is necessary. In order to address these challenges, genetic algorithms have been developed for the calibration of high channel count OPAs as well as an advanced characterization setup compatible with wafer-scale OPA characterization.

## ACKNOWLEDGEMENTS

This work was partially funded by the French ANR via Carnot funding, the ECSEL Vizta European project and the French National Program "Programme d'investissement d'avenir, IRT Nanoelec, n° ANR-10-AIRT-05".